\documentclass[a4paper]{jpconf}
\usepackage{graphicx}
\usepackage{amsmath,amssymb,amsfonts,amsthm,latexsym} 
\usepackage[pdftex,colorlinks=true,urlcolor=blue,linkcolor=blue,citecolor=blue,a4paper]{hyperref}
\usepackage{color}
\usepackage{float}
\begin{document}

\title{Radio astronomy in Africa: the case of Ghana}

\author{B D Asabere$^1$, M J Gaylard$^2$, C Horellou$^3$, H Winkler$^1$ and T Jarrett$^4$ }

\address{$^1$ Department of Physics, University of Johannesburg, P.O. Box 524, 2006, Auckland Park, Johannesburg, South Africa}
\address{$^2$ HartRAO, P. O. Box 443, Krugersdorp, South Africa (passed away on 14 Aug 2014)}
\address{$^3$ Department of Earth and  Space Sciences, Chalmers University of Technology, Onsala Space Observatory, SE-439 92 Onsala, Sweden}
\address{$^4$ Department of Astronomy, University of Cape Town,  Private Bag X3,  Rondebosch 7701, South Africa}

\ead{ bd.asabere@gmail.com}

\begin{abstract}
South Africa has played a leading role in radio astronomy in Africa with the Hartebeesthoek Radio Astronomy Observatory (HartRAO). It continues to make strides with the current seven-dish MeerKAT precursor array (KAT-7), leading to the  64-dish MeerKAT and the giant Square Kilometer Array (SKA), which will be used for transformational radio astronomy research. Ghana, an African partner to the SKA, has been mentored by South Africa over the past six years and will soon emerge in the field of radio astronomy. The country will soon have a
science-quality 32m dish converted from a redundant satellite communication antenna. Initially, 
it will be fitted with 5 GHz and 6.7 GHz receivers to be followed later by a 1.4 - 1.7 GHz
receiver. The telescope is being designed for use as a single dish observatory and for participation in the
developing African Very Long Baseline Interferometry (VLBI) Network (AVN) and the European VLBI
Network. Ghana is earmarked to host a remote station during a possible SKA Phase 2. The country's location of 5$^\circ$ north of the Equator gives it the distinct advantage of viewing the entire plane of the Milky Way galaxy and nearly the whole sky. In this article, we present Ghana's story in the radio astronomy scene and the science/technology that will soon be carried out by engineers and astronomers.

\end{abstract}

\section{Introduction}
\label{Introd}  
In the field of radio astronomy, South Africa (SA) has been the pacesetter on the African continent, 
with the long established Hartebeesthoek Radio Astronomy Observatory (HartRAO). It has world-class astronomical and space research facilities for cutting-edge radio astronomy research and studies. The current seven-dish  MeerKAT precursor array (KAT-7), leading to the 64-dish MeerKAT with first light in late 2016 and the giant
Square Kilometer Array (SKA) which construction will start in 2018, will be used for unprecedented radio
astronomy researches that will lead to new discoveries. In 1992, Mauritius also appeared on the scene with its meter-wave Fourier Synthesis T-shaped array, the Mauritius Radio Telescope (MRT), designed to survey the southern sky for point sources at 151.6~MHz in the declination of -70$^\circ$ to -10$^\circ$ and sensitivity of 200mJy \cite{Smanah2002}. The MRT survey produced a southern sky equivalent of the Sixth Cambridge Catalog (6C) of bright radio sources \cite{Baldwin1985}. Although several other African countries had nurtured strategies and plans to host
radio astronomy facilities, they are yet to come to fruition. Attempts by Nigeria to build a HartRAO-like facility at Nsukka \cite{Okere2011} and Egypt's plans \cite{Shaltout1999, Mosalam1999} to operate a radio telescope, in the frequency range of 1.4 to 43~GHz, at Abu Simbel in the southern part of the country are all yet to succeed.
\medskip

In the mid-2000's, the call on Botswana, Ghana, Kenya, Madagascar, Mauritius, Mozambique, Namibia and Zambia to partner South Africa in its bid to host the SKA, produced new motivation for countries on the continent to speed up their efforts in embracing radio astronomy studies and researches. With the KAT-7 successes \cite{Woudt2013}, the 64-dish MeerKAT project gearing up and the decision that Africa would host a large part of the SKA, the South African SKA Project (SKA-SA) and HartRAO searched in the partner countries to identify redundant satellite communication antennas with potential for conversion of such expensive, but now obsolete, assets for radio astronomy. This type of conversion (e.g. \cite{Fujisawa2002, McCulloch2005, Mansfield2010}) has been made possible globally owing to the switching over from the data streaming communication satellites to the more efficient and cheaper undersea fibre optic cables for telecommunication signal transport. The goal of the exercise, which led to the discovery of many 30m-class antennas across Africa, is to build the needed capacity in support staff, engineers and scientists in radio astronomy and related disciplines. Ghana has such an asset at its Kuntunse Intelsat Satellite Communication Earth Station (see figure~\ref{dish_cables} {\it Left} panel)

\medskip
In this paper, we give a short briefing on Ghana's radio astronomy prospects and outline the possible science/technology that will soon be done from there. 

\section{The Ghana Antenna}
\label{sect2}  
The Ghana Intelsat Satellite Earth Station at Kuntunse is situated at an elevation of 70m above sea level with position coordinates of 05$^\circ$ 45$'$ 01.5$''$ N  and 00$^\circ$ 18$'$ 18.4$''$ W \cite{Ayer2008}. It hosts three antennas of diameters 32m, 16m and 9m, but only the 16m antenna is still operational for satellite communication. The station was commissioned on August 12, 1981  and was operated by the Ghana Telecommunications Corporation until July 03, 2008 when Ghana Vodafone took over, as major shareholder (i.e. purchased 70\% shares). Kuntunse is a surburb located off the Nsawam Road, about 25km north-west of the national capital, Accra.

\begin{figure}[!tb]  
\begin{minipage}{13pc}
\includegraphics[width=7.0cm,height=7.8cm]{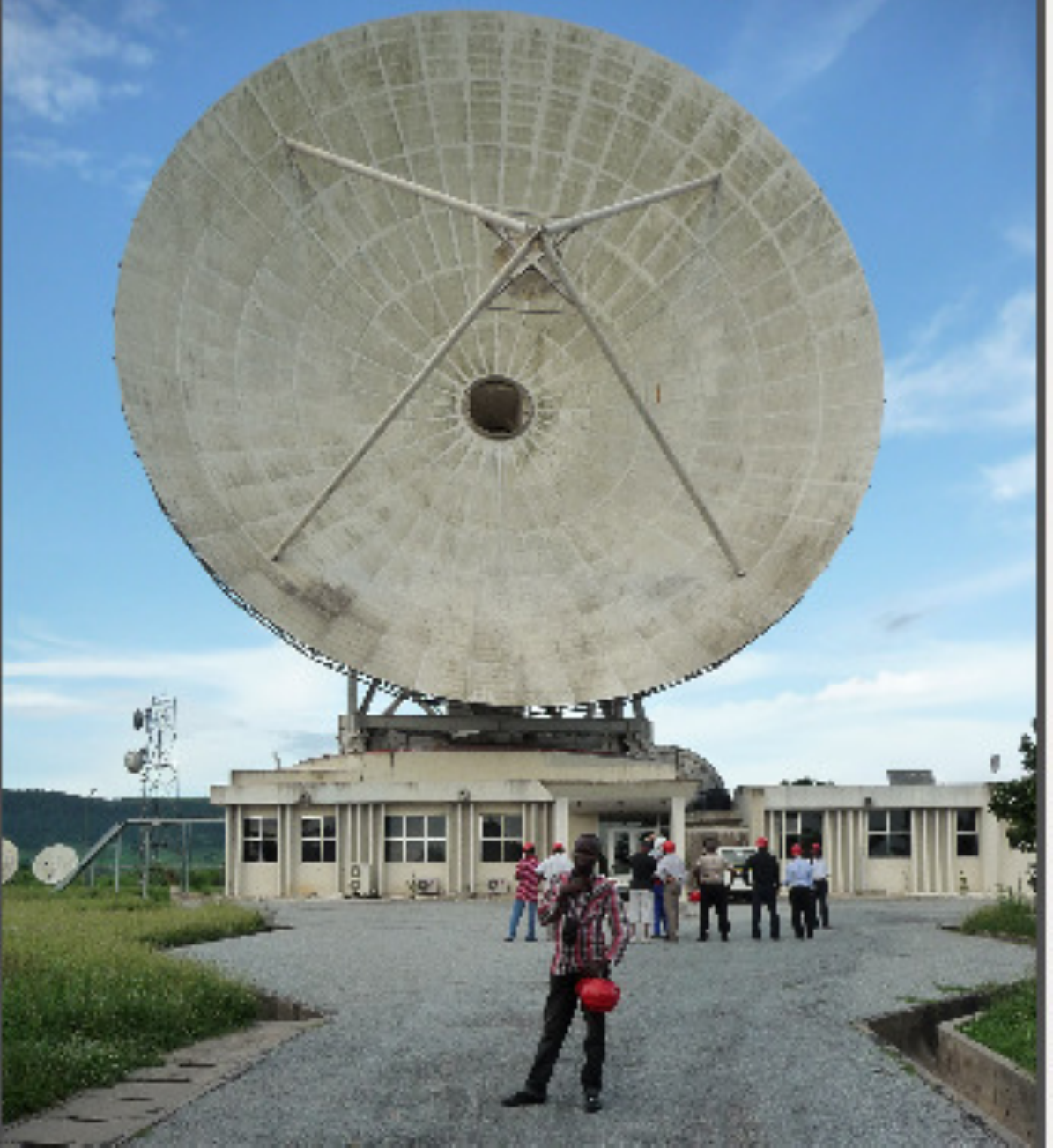}
\end{minipage}\hspace{6pc}%
\begin{minipage}{13pc}
\includegraphics[width=7.4cm,height=7.8cm]{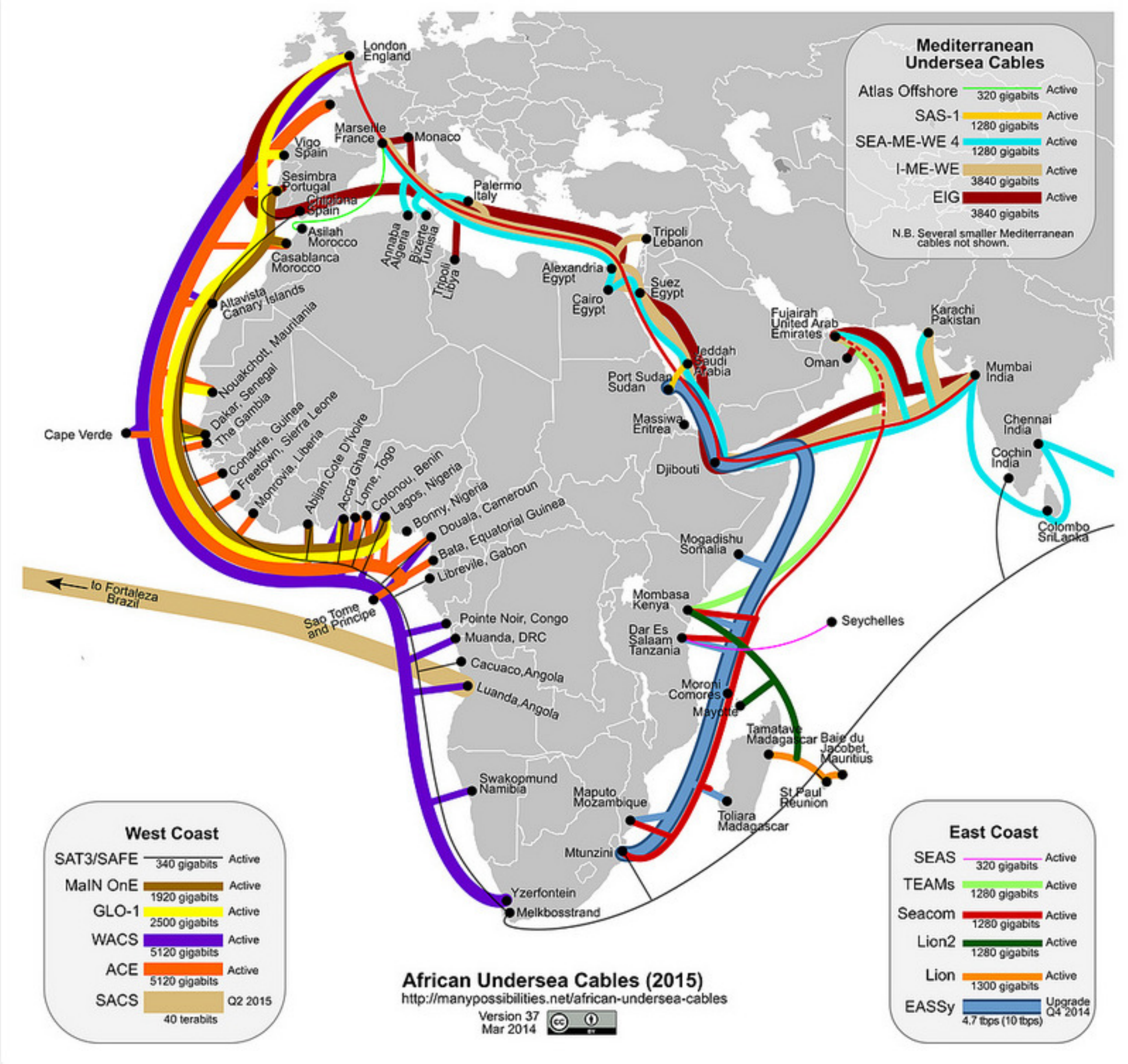}
\end{minipage}\hspace{6pc}%
\caption{{\it Left:} The Kuntunse 32m disuse telecommunication antenna currently being converted to radio astronomy facility. Standing in front of the telescope is the first author. 
{\it Right:} African Undersea Cables span the coasts of Africa and link the continent to the outside World and represents greater internet bandwidth availability and potential of data transport from the Ghana antenna site  \cite{Underseacables}.}
\label{dish_cables}  
\end{figure}

\medskip  
Ghana has warmly embraced the strategy into radio astronomy, according to its radio astronomy development plan incorporated in {\it The Ghana Science, Technology and Innovation Policy, and The Science, Technology and Innovation Development Plan 2011-2015 (GPDP15)}. As an SKA Africa partner country, it welcomed and collaborated with the SKA-SA/HartRAO group to access the radio astronomy potential of the redundant satellite communication antennas at Kuntunse. The suitability of the 32m cassegrain antenna and the Kuntunse control station (see figure~\ref{dish_cables} {\it Left} panel) for radio astronomy were established by the group after two successive working visits in March and May 2011.  

\subsection{The Antenna Conversion}
To facilitate the agenda envisaged in the GPDP15, the Government under the Ministry of Environment, Science, Technology and Innovation (MESTI) on January 01, 2011 established an institution called the  Ghana Space Science and Technology Institute (GSSTI), to spearhead all radio astronomy, space science and emerging related technologies programmes, activities and research. GSSTI was established under the  Ghana Atomic Energy Commission (GAEC), a national research organisation. It started initially as a Centre under GAEC's Graduate School of Nuclear and Allied Sciences (SNAS), an affiliation of the University of Ghana.

\medskip
After some protracted negotiations and the official handing over of the station to the state, the antenna conversion exercise by a team of scientists and engineers from SKA-SA/HartRAO and GSSTI started in earnest. Besides the astronomy instrumentation upgrade, major rehabilitative work includes replacing the corroded subreflector quadrupod support legs, replacing the azimuth and elevation angle resolvers with more accurate angle encoders, covering the beam waveguide aperture with radome, replacing some rusty hardware on the antenna structure, replacing azimuth and elevation limit switches, flushing and greasing the gearbox systems, changing the elevation and azimuth oil gauges, repainting the whole antenna structure and re-engineering the automatic control/rotation system. The exercise has been running parallel with the requisite local human capital development (HCD) to produce the needed manpower to be custodians and users of the facilities via both formal and informal training interventions. A team of graduates from GSSTI has just completed 6 months training with the South African AVN team at SKA-SA and HartRAO. The latest target for test observations with the 32m antenna is June 2015, with science operations targeted for the end of 2015.

\subsection{The Antenna Receivers}
The type of observations and science that can best be done with a radio telescope depend on the telescope's location, size, specifications and the receivers and science instruments it is fitted with. During the testing period (Phase-1), the existing telecommunication feed horn in the frequency range 3.8 - 6.4~GHz (C-band) will be used. For the actual science observations (Phase-2), it will initially be fitted with uncooled 5 GHz and 6.7 GHz (C-band) receivers  to be followed later by a 1.4 - 1.7 GHz or wider L-band receiver, for which extra funding will be needed. Future receiver developments could include replacing the original C-band feed horn with a wider band design covering more VLBI bands, introducing cryogenic receivers for improved sensitivity and adding more frequency bands.

\medskip
\subsection{Funding for the Conversion}
The funds for the antenna conversion and HCD has so far come from the African Renaissance Fund (ARF) of South Africa's Department of International Relations and Cooperation (DIRCO), South Africa's Department of Science and Technology (DST) and the Government of Ghana. The SKA-SA Project, HartRAO and GSSTI are the main facilitating and implementing bodies for the Ghana conversion.

\section{Radio Astronomy Projects}
\label{sect3}  
On realising a full operational 32m radio astronomy telescope, there are a number of planned astronomical projects: 

\begin{itemize}
 \item Operate as a single dish observatory, to be known as the Ghana Radio Astronomy Facility (GRAF).
  \item For participation in the developing 4-dish African VLBI Network (AVN) and possible expanded future AVN \cite{Gaylard2011}. 
   \item For taking part in the European VLBI Network (EVN).
   \item With antennas east of South Africa and central Africa, the Ghana antenna would be relevant to both the Australian and American VLBI Arrays. Again, the Ghana antenna, and other AVN telescopes, would be valuable in adding long baseline capability to SKA Phase 1.
  \item For the possible event of SKA Phase 2, Ghana is earmarked to host some of the single-pixel feed dishes as a remote station. 
\end{itemize}

 \begin{figure}[!tb] 
\begin{minipage}{13pc}
\includegraphics[width=6.8cm,angle=-90]{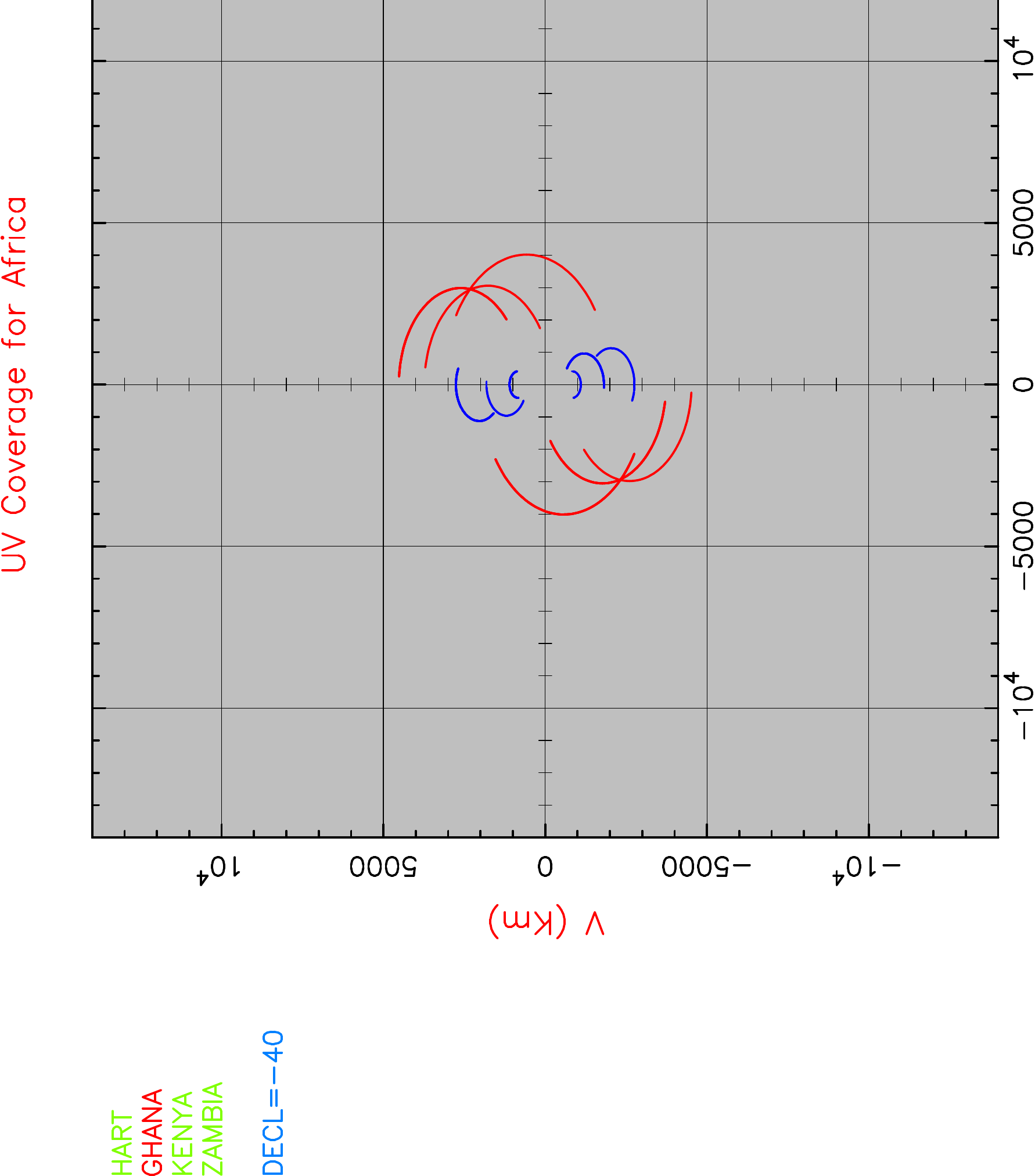}
\end{minipage}\hspace{6pc}%
\begin{minipage}{13pc}
\includegraphics[width=6.8cm,angle=-90]{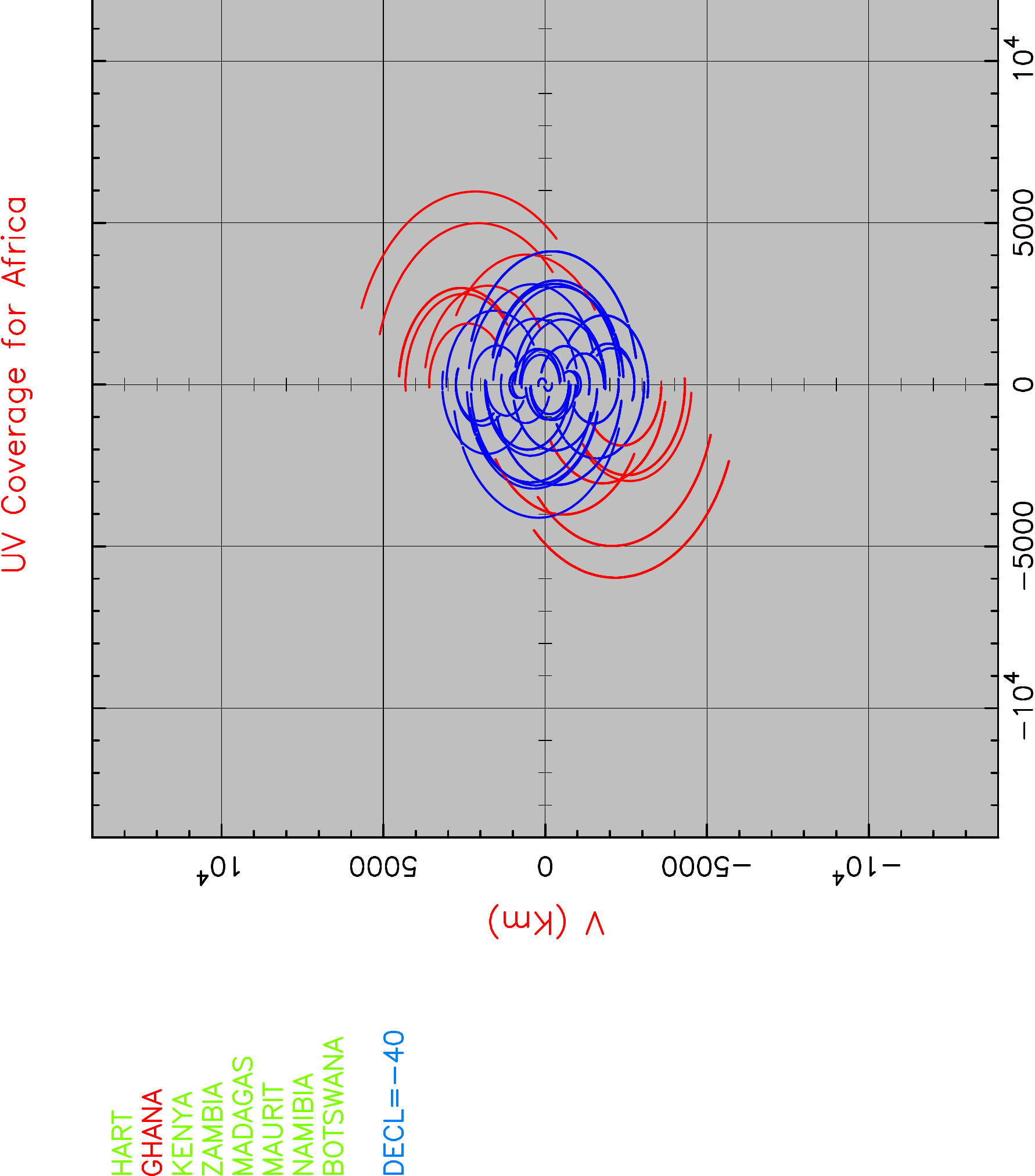}
\end{minipage}\hspace{6pc}%
\caption{VLBI UV coverages of the developing AVN showing the impacts (in red tracks) of the Ghana 32m antenna in: {\it Left:} A 4-dish developing AVN, involving the antennas in Ghana (32m), HartRAO (26m), Kenya (32m) and Zambia (30m or 32m), tracking a source  at $-$40$^\circ$ declination. {\it Right:} A network of antennas in seven SKA-SA partner countries, together with HartRAO, tracking a source at $-$40$^\circ$ declination.}
\label{africavlbi}   
\end{figure}

\section{Science with the Ghana Antenna}
\label{sect4} 
Ghana's location of 5$^\circ$ north of the Equator (see Section~\ref{sect2}) gives it the advantage of viewing the entire plane of the Milky Way galaxy and nearly the whole sky (see figure~\ref{africavlbi}). Another competitive advantage for Ghana, is that it is located close to the African Undersea Cables \cite{Underseacables} (see {\it Right} panel of figure~\ref{dish_cables}) spanning the east and west coasts of the continent and linking Africa to the rest of the World. This close proximity promises greater bandwidth and faster internet connectivity for data transport. With such valuable resources much can be done from Ghana. 

\medskip
In this section we describe some of the science cases of the Kuntunse telescope as a single dish and in VLBI networks, and for the possible long baseline array of SKA Phase 2.


\subsection{Single-Dish Science Cases}
With the C- and L-bands receivers fitted for Phase-2 operations, one can do the following with the Ghana antenna:
\begin{itemize}
\item Radio Continuum Flux measurements (with wideband multi-channel radiometer); use known radio astronomy calibration sources for daily calibrations of receivers and also follow radio emissions from sources such as AGNs emitting gamma-ray flares.
\item Pulsar Observations (with wideband multi-channel pulsar timer); monitor the behaviour of pulsars of interest over a long period of time, such as those producing glitches and intermittent pulsars, and hunt for fast radio burst sources. 
\item Emission Lines Spectroscopy (with narrowband multi-channel spectrometer); maser line monitoring of star forming regions, including hydroxyl masers (1612, 1665, 1667, 1720 MHz) and methanol masers (6668 MHz). 
\end{itemize}

\subsection{VLBI Networks Science Cases} 
\begin{figure}[!tb]  
\begin{minipage}{13pc}
\includegraphics[width=6.8cm,angle=-90]{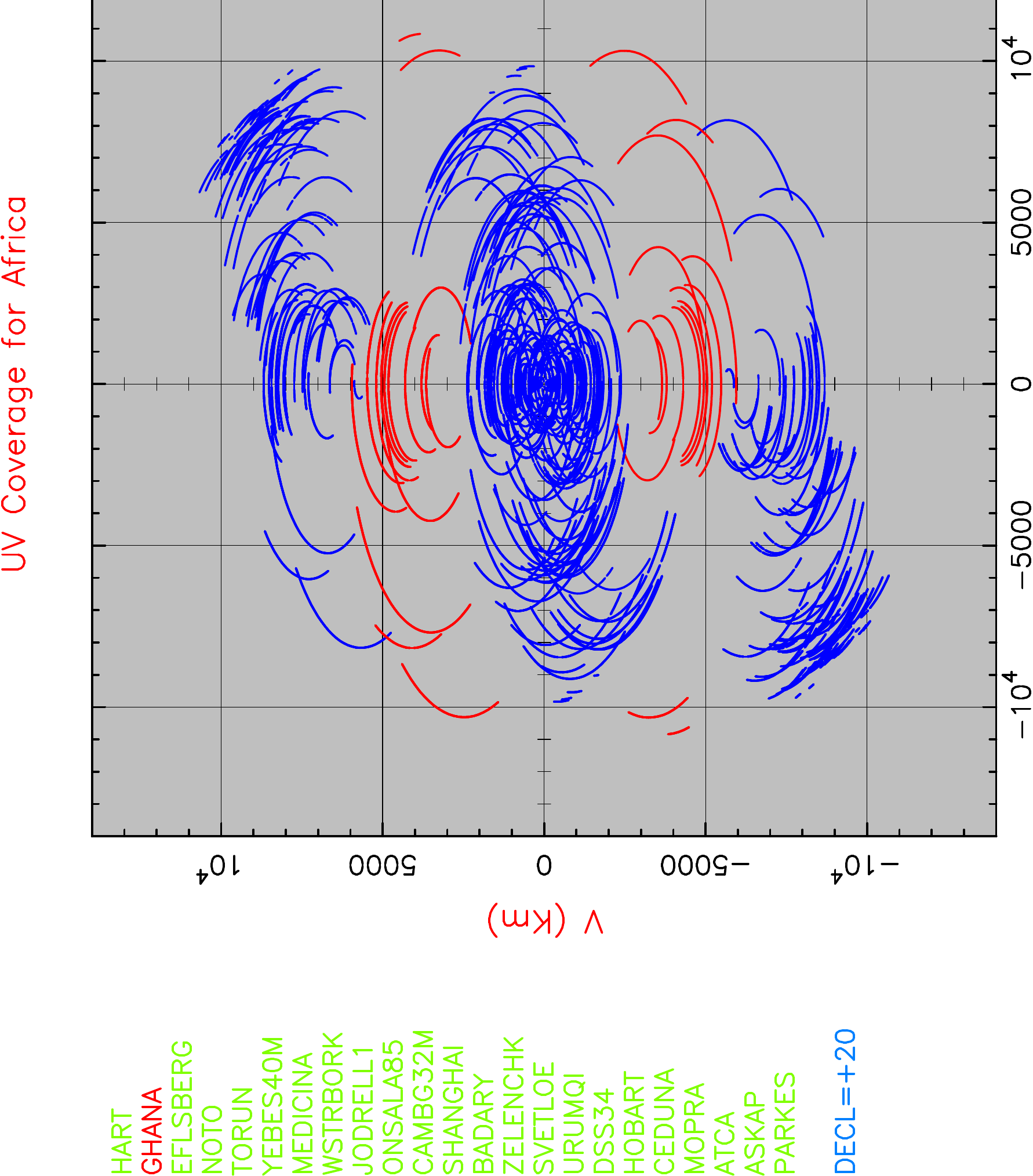}
\end{minipage}\hspace{6pc}%
\begin{minipage}{13pc}
\includegraphics[width=6.8cm,angle=-90]{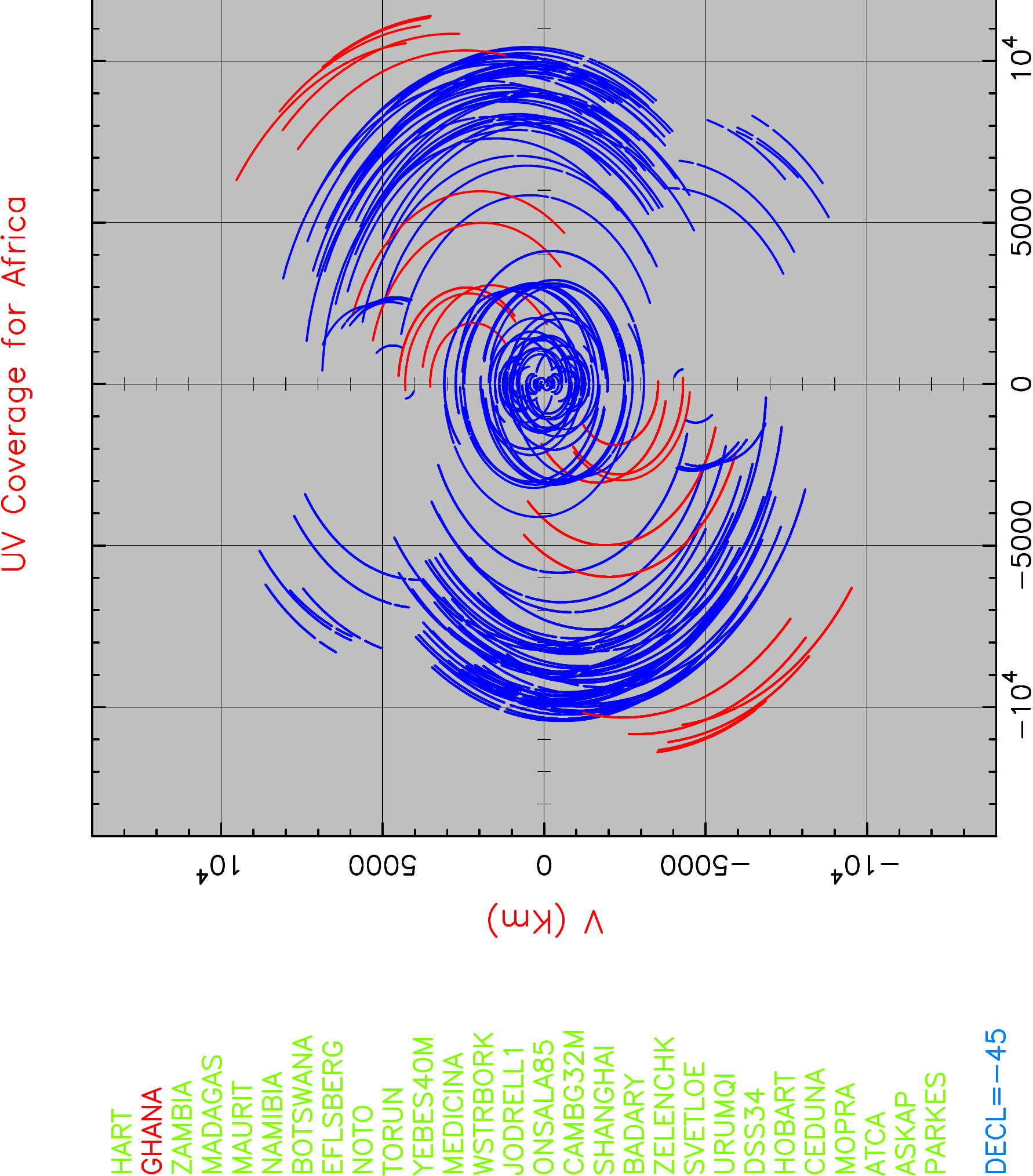}
\end{minipage}\hspace{6pc}%
\caption{Global VLBI UV coverages of full-track observations showing input from the Ghana 32m antenna in red tracks observing a source: {\it Left:} At the declination $+$20$^\circ$  with the existing VLBI antennas.
{\it Right:} At the $-$45$^\circ$ declination with other five proposed AVN antennas in Africa.}

\label{globalvlbi}  
\end{figure}

With the System Equivalent Flux Density (SEFD) of the dish designed to be better than the typical threshold (i.e. SEFD $<$ 800Jy) for telescopes in current VLBI networks, the Ghana antenna can be a valuable part of the existing VLBI networks. In both the stand-alone AVN, EVN and global VLBI, the inclusion of Ghana will improve imaging and calibration quality and sensitivity in all VLBI astronomy science cases (refer to figures~\ref{africavlbi} and ~\ref{globalvlbi}). The VLBI science cases that will be enhanced with the addition of the Ghana telescope equipped for C- and L-band include:

\begin{itemize}
\item Mapping Interstellar masers in star-forming regions in the Milky Way.
\item Determining the distances to star-forming regions in the Milky Way through methanol maser parallax measurement.
\item Using trigonometric parallax measurements to determine accurate pulsar distances as well
as pulsar proper motions.
\item Imaging active galactic nuclei (AGN).
\item Resolving binary systems in extragalactic supermassive black holes.
\item Searching for radio transients - long baselines provide discrimination against radio frequency interference.
\item Imaging radio emission from X-Ray binary systems and relativistic jets.
\end{itemize}

\subsection{Science at Very High Angular Resolution} 
In the likely event of Phase 2 of the SKA, Ghana would host a 30-dish single-feed antenna array on the proposed 400m$\times$400m piece of land in the Kwahu mountain range area located in the south-central eastern corridor of the country. This will be in support of the SKA goals of doing science at very high angular resolution (see \cite{Godfrey2012}).

\section{Other Relevant Science Instruments} 
\label{sect5} 
With broad bandwidth and fast internet facilities' availability at the Kuntunse Satellite Earth Station, other instruments such as Meteorological unit (MET-4), Global Navigational Satellite System (GNSS) Receivers, Seismometer, Gravimeters and Magnetometers may be cost-effectively added to enhance the science capability of the station for the benefits of the global community.

\section{Summary}
We have unveiled a strong case for the conversion of an obsolete telecommunication facility for radio astronomy use, which presents an excellent prospect of doing radio astronomical science from Ghana. The science cases could be more or less than those outlined in the paper, based on availability of funding and partnerships. The custodians of the dish, GSSTI, warmly welcomes partner support of all forms to realize a befitting facility for radio astronomy and technological science use.

\section*{Acknowledgement}
The first author acknowledges funding from MIDPREP, an exchange programme between two European  Institutes (Chalmers and ASTRON) and three South African partners (University of Stellenbosch, Rhodes University and University of Cape Town). He expresses profound appreciation to SKA-SA/HartRAO and the staff at the Onsala Space Observatory in Sweden for assistance during his regular visits, to mention Jun Jang of the VLBI group for the help.


\end{document}